\documentclass[11pt]{article}
\usepackage{epsfig}
\usepackage{graphicx}
\usepackage{amsmath}
\usepackage{amssymb}
\usepackage{amsxtra}
\usepackage{amsfonts,amsthm}
\usepackage{subfigure}
\usepackage[rflt]{floatflt}
\usepackage{color}
\usepackage[T1]{fontenc}
\usepackage[latin9]{inputenc}
\def\Journal#1#2#3#4{{#1} {\bf #2} (#4) #3 }
\def\RNC{ Rivista Nuovo Cimento}

\def\PRL{ Phys. Rev. Lett.}

\def\BWP{ Bled Workshops in Physics}

\def\lsim{\mathrel{\rlap{\lower4pt\hbox{\hskip1pt$\sim$}}
    \raise1pt\hbox{$<$}}}         %less than or approx. symbol

\def\gsim{\mathrel{\rlap{\lower4pt\hbox{\hskip1pt$\sim$}}
    \raise1pt\hbox{$>$}}}         %greater than or approx. symbol
\title{The nuclear physics of OHe }
\author{J.R. Cudell$^{a,}$\footnote{JR.Cudell@ulg.ac.be}\\
M.~Khlopov $^{b,}$\footnote{Khlopov@apc.univ-paris7.fr}\\
Q. Wallemacq
$^{a,}$\footnote{Quentin.Wallemacq@ulg.ac.be}}
\begin{document}
%%%%%%%%%%%%%%%%%%%%%%%%%%%%Title page
\maketitle
{{\it $^{a}$ \small IFPA, D\'ep. AGO, Universit\'e de Li\`ege, Sart Tilman, 4000 Li\`ege, Belgium\\\
      $^{b}$ \small National Research Nuclear University ``Moscow Engineering Physics
Institute'', 115409 Moscow, Russia \&
Centre for Cosmoparticle Physics ``Cosmion'' 115409 Moscow, Russia
\& APC laboratory 10, rue Alice Domon et Lonie Duquet
75205 Paris Cedex 13, France}}

\begin{abstract}
A recent composite-dark-matter scenario assumes that the dominant 
fraction of dark matter consists of O-helium (OHe) dark atoms, in which a lepton-like 
doubly charged particle O$^{--}$ is bound with a primordial helium nucleus. It liberates 
the physics of dark matter from unknown features of new physics, but it demands
a deep understanding of the details of known nuclear and atomic physics, which are still 
unclear. Here, we consider in detail the physics of the binding of
OHe to various nuclei of interest for direct dark matter searches. We show that
standard quantum mechanics leads to bound states in the keV region, but does
not seem to provide a simple mechanism that stabilizes them. The crucial role of a 
barrier in the OHe-nucleus potential is confirmed for such a stabilization.
\end{abstract}

\section{Introduction}
Direct searches for dark matter have produced surprising results. Since the DAMA collaboration observed a
signal, several other collaborations seem to confirm an observation, while others
clearly rule
out any detection. We summarize the situation in Table \ref{searches}, and the current
experimental situation is reviewed in \cite{DAMAreview}. This apparent
contradiction comes from the analysis of the data under the assumption that nuclear
recoil is the source of the signal.

Starting from 2006 it was proposed \cite{invention,spectro} that the
signal may be due to a different source: if dark matter has weakly bound states with
normal matter, the observations could come from radiative capture of thermalized dark
matter, and could depend on the detector composition and temperature. This scenario comes
naturally from the consideration of composite dark matter. Indeed, one can imagine that
dark matter is the result of the existence
of heavy negatively charged particles that bind to primordial nuclei.

Cosmological considerations imply that such candidates for dark matter should consist of
negatively doubly-charged heavy ($\sim 1$ TeV) particles, which we call O$^{--}$, coupled
to primordial helium. Lepton-like technibaryons, technileptons, AC-leptons or clusters of
three heavy anti-U-quarks of 4th or 5th generation with strongly suppressed hadronic
interactions are examples of such O$^{--}$ particles (see \cite{invention,spectro} for
a review and for references).

The cosmological and astrophysical effects of such composite dark matter (dark atoms of
OHe) are dominantly related to the helium shell of OHe and involve only one parameter
of new physics $-$ the mass of O$^{--}$. The positive results of the DAMA/NaI and
DAMA/LIBRA
experiments are explained by annual modulations of the rate of radiative capture of OHe
by sodium nuclei. Such radiative capture is possible only for intermediate-mass nuclei:
this explains the negative results of the XENON100 experiment. The rate of this capture
is
proportional to the temperature: this leads to a suppression of this effect in cryogenic
detectors, such as CDMS. OHe collisions in the central part of the Galaxy lead to OHe
excitations, and de-excitations with pair production in E0 transitions can explain the
excess of the positron-annihilation line, observed by INTEGRAL in the galactic bulge.

These astroparticle data can be fitted, avoiding many astrophysical uncertainties of WIMP 
models, for a mass of O$^{--}$ $\sim 1$ TeV, which stimulates searches for stable doubly 
charged lepton-like particles at the LHC as a test of the composite-dark-matter scenario.
%%%%%%%%%%%%%%%%%%%%%%%%%%%%%%%%%%%%%%%%%%%%%%%%%%%%%%%%%%%%%%%%%%%%%%%%%%%%%%%%%%%%%%%%%%%%%%%
\begin{table}
\begin{center}
\begin{tabular}{|c|c|c|c|c|c|}
\hline
Detector  & nuclei & A&Z& temperature&detection \\
\hline
DAMA       &Na & 23      &11   &300 K&8.9 $\sigma$\\
(/NaI \cite{DAMA}&I&127&53&&\\
+/LIBRA \cite{DAMALIBRA})&Tl&205&81&&\\ \hline
CoGeNT\cite{COGENT}         &Ge     & 70-74         &32       &70 K&2.8 $\sigma$\\ \hline
CDMS\cite{CDMS}             & Ge&  70-74 & 32  &cryogenic& $-$\\
             &(Si) & (28-30)  &(14)   && \\\hline
%CRESST\cite{CRESST}         & O Ca W& 16, 40-44, 182-186 &8, 20, 74  &20 mK&4.2
%$\sigma$\\
%EDELWEISS\cite{EDELWEISS}   &Ge     & 70-74&32   & &$-$\\
XENON100\cite{XENON}        &Xe     & 124-134        &54 & cryogenic   &$-$\\
\hline
\end{tabular}
\caption{Results of various dark matter searches and composition of the detectors.
}\label{searches}
\end{center}
\end{table}
%%%%%%%%%%%%%%%%%%%%%%%%%%%%%%%%%%%%%%%%%%%%%%%%%%%%%%%%%%%%%%%%%%%%%%%%%%%%%%%%%%%%%%%%%%%%
The problem with OHe dark matter is that its constituents may interact too much
with normal matter. OHe is neutral, but a priori it has an unshielded nuclear attraction 
to matter nuclei. To avoid the problem, it was assumed that the effective
potential between OHe and a normal nucleus would have a barrier, preventing He and/or
O$^{--}$ from falling
into the nucleus, allowing only one bound state, and diminishing considerably the
interactions of OHe. Under these conditions elastic collisions dominate in OHe
interactions with matter, which is important for many aspects of the OHe scenario.

In this paper, we show that indeed such a barrier is needed to make the model work, and
we try to establish its existence through several methods. In
the first section, we review the classical description of the problem \cite{spectro} and 
show that in fact
it does
not lead to a repulsive force. In section 2, we explore the spectrum of the bound states
of OHe. We show that, if one considers only the screened Coulomb force, then bound states
exist only for light nuclei, whereas if we consider a polarization of OHe due to a
second-order Stark effect, then most nuclei have keV bound states.
In the last section, we
check that the description of the Stark effect that we used is reasonable via a
perturbative calculation at large distances, but it is not reliable when the 
nucleus comes close to OHe, as one would then need to take into account
a strong and inhomogeneous deformation of the ground 
state by the common effect of Coulomb and nuclear force.

\section{Classical model}
\begin{figure}
\begin{center}
\includegraphics[scale=0.25]{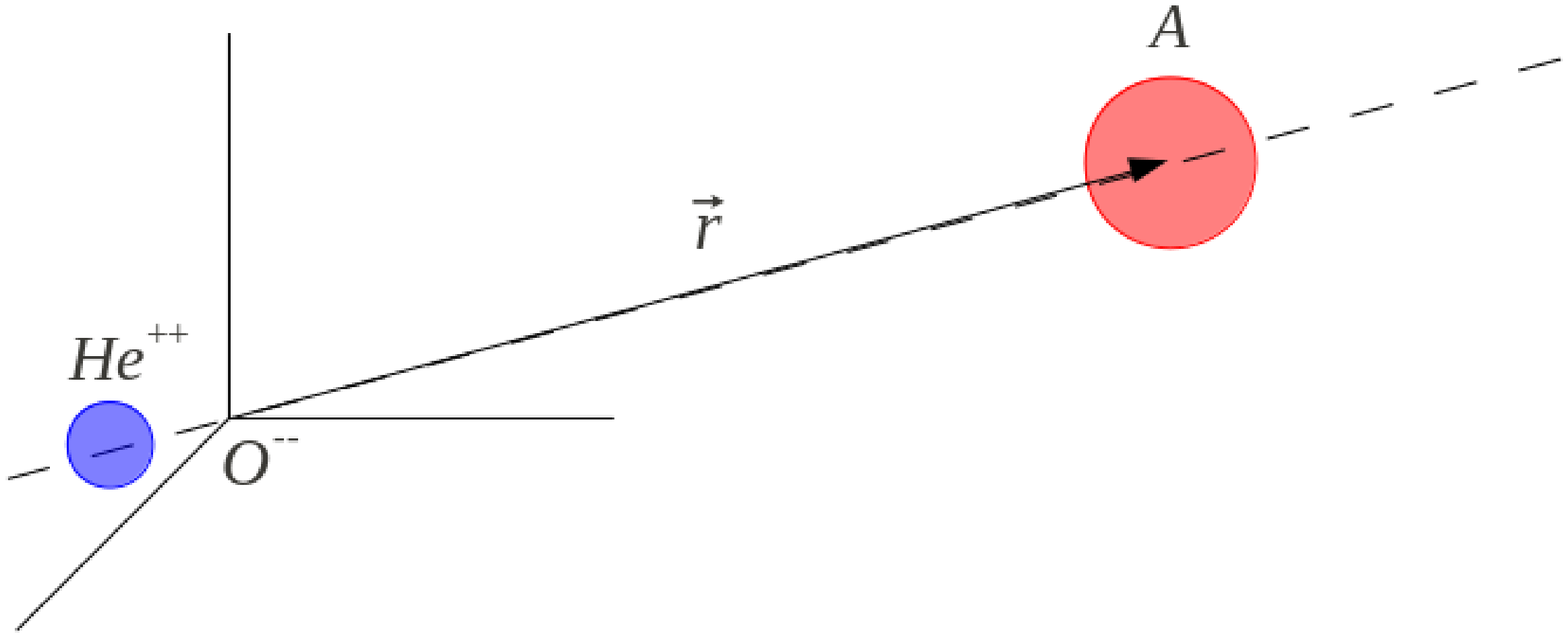} \\
\includegraphics[scale=0.25]{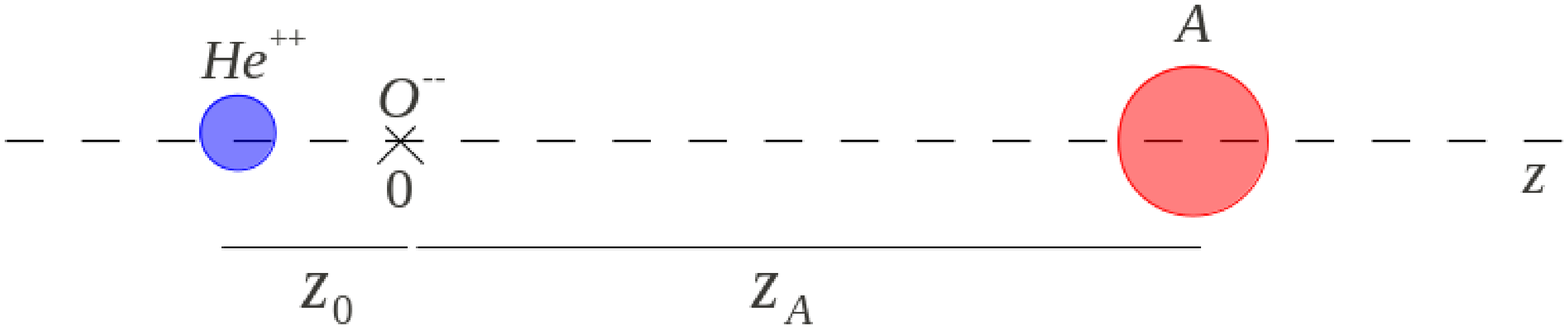}
\end{center}
\caption{One-dimensional OHe atom polarized along the O$^{--}-$A axis, denoted
$z$. $z_{0}$ is the distance between O and He and $z_{A}$ is the
distance of the A nucleus along the $z-$axis.\label{fig:One-dimensional-OHe-atom}}
\end{figure}

To study the polarization of the OHe atom under the influence of an
approaching A nucleus, we can first treat OHe as a classical
structure and neglect the effects of O$^{--}$ and nucleus motion. 
The polarization
of OHe is then fixed by the equilibrium of forces
acting on the He nucleus. For every position of the A nucleus,
we can work on the O-A axis, in the rest frame of the O$^{--}$
particle, as shown in Fig.~\ref{fig:One-dimensional-OHe-atom}.

We take the He and A nuclei as uniformly charged spheres of radii $R_{He}$
and $R_{A}$ and of charges $Z_{He}=2$ and
$Z_{A}$. We also assume that O$^{--}$ is point-like. We then obtain the electrostatic potential for the interactions with O$^{--}$ :
\begin{eqnarray}
V_{OA}(z_{A}) & = & -\frac{Z_{A}Z_{O}\alpha}{z_{A}}, {\rm\ for\ } z_{A}>R_{A}\nonumber\\
 & = & \frac{-Z_{A}Z_{O}\alpha}{2R_{A}}\left(3-\frac{z_{A}^{2}}{R_{A}^{2}}\right),{\rm\ for\ } z_{A}<R_{A}\label{eq:2}\end{eqnarray}
for the interaction between the O$^{--}$ and the A nucleus, and a similar expression for $V_{OHe}(z_{He})$.
The potential between the two nuclei has both electrostatic and nuclear contributions.
In the former, we neglect the He size, and for the latter we use an experimental parametrisation of the $\alpha$-nucleus potential from scattering experiments \cite{key-1}:
\begin{eqnarray}
V_{HeA}(z_{A}-z_{0}) & = & \frac{Z_{He}Z_{A}\alpha}{|z_{A}-z_{0}|}+\frac{-V_{0}}{1+e^{\left(|z_{A}-z_{0}|-R_{*}\right)/a}},{\rm\ for\ }|z_{A}-z_{0}|>R_{A}\nonumber\\
 & = & \frac{Z_{He}Z_{A}\alpha}{2R_{A}}\left(3-\frac{\left(z_{A}-z_{0}\right)^{2}}{R_{A}^{2}}\right)+\frac{-V_{0}}{1+e^{\left(|z_{A}-z_{0}|-R_{*}\right)/a}},\nonumber\\
 && \ \ \ \ \ \ \ \ {\rm\ for\ } |z_{A}-z_{0}|<R_{A}.\label{eq:3}\end{eqnarray}
The nuclear
interaction is represented in a Woods-Saxon form, with parameters $V_{0}=30$ MeV,
$a=0.5$ fm and $R_{*}(fm)=1.35\times A^{1/3}+1.3$ (fm).

The equilibrium position $z_{0}$ of the He nucleus will be at the minimum of the potential $V_{HeA}+V_{OHe}$ and will depend on $z_{A}$. At that point, the Coulomb
force balances the nuclear force:
\begin{equation}
\overrightarrow{F}_{OHe}+\overrightarrow{F}_{HeA}=\overrightarrow{0}.\label{eq:4b}\end{equation}
When the equilibrium position is determined, the OHe-A potential
is obtained by adding the dipole potential to the Woods-Saxon one \begin{equation}
V_{OHeA}(z_{A})=V_{dip}(z_{A})+V_{WS}(z_{A}-z_{0}),\label{eq:5}\end{equation}
 where \begin{equation}
V_{dip}(z_{A})=\frac{2Z_{A}\alpha z_{0}(z_{A})}{z_{A}\left(z_{A}-z_{0}(z_{A})\right)}\label{eq:6}\end{equation}
 is the dipole potential of the polarized OHe atom.

\begin{figure}
\begin{centering}
\includegraphics[scale=0.7]{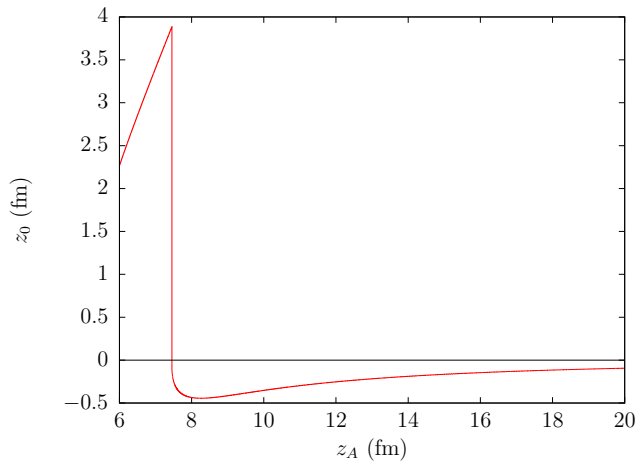}
\par\end{centering}
\caption{Polarization $z_{0}$ of the one-dimensional OHe atom as a function
of the distance $z_{A}$ of an approaching sodium nucleus. The break in the curve
corresponds to the fact that for a distance of about 9 fm,
He falls into the nuclear potential of A. \label{fig:Polarization--of}}
\end{figure}

Fig.~\ref{fig:Polarization--of} shows the polarization $z_{0}$
of the OHe atom as a function of the position $z_{A}$
along the $z-$axis for an approaching sodium nucleus with $Z_{A}=11$.
We see that it is negative at large distance, giving rise to an
attractive dipole potential and that the nuclear force starts to reverse
the dipole when the nucleus gets closer to O$^{--}$.

\begin{figure}
\begin{centering}
\includegraphics[scale=0.7]{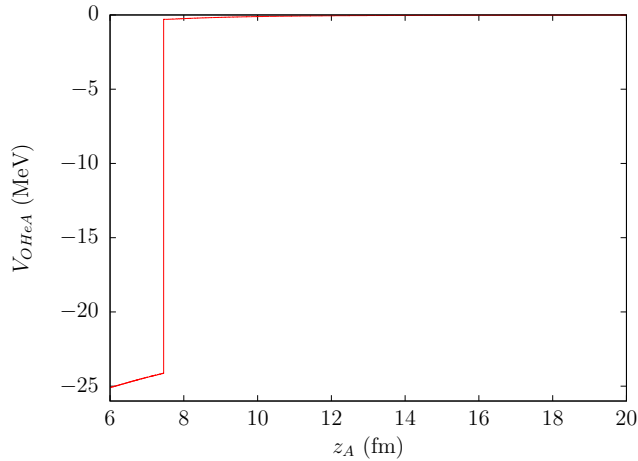}
\par\end{centering}
\caption{Total OHe-A potential for sodium.\label{fig:Total-OHe--potential}}
\end{figure}

This situation
corresponds to a repulsive dipole that acts against the nuclear force,
but it can be seen in
Fig.~\ref{fig:Total-OHe--potential}
that this repulsive force
is not sufficient to overcome the nuclear force between the two nuclei
and to give rise to a repulsive global potential.

When $z_{A}\lesssim7.5$ fm,
Equation \eqref{eq:4b} projected along the $z-$axis loses its initial
solution and another one remains, that is located in the nuclear well,
giving rise to a jump in the polarization and therefore in the total
potential. Similar results and pictures can be obtained for other
nuclei.

Hence we see that, classically,
 no repulsive potential appears, even if the electrostatic force of OHe repels
 the A nucleus. In fact one can argue that this is a generic classical result
 which does not depend on the details of the calculation. If the configuration
 of the 3 objects is He-O-A, then clearly the force is attractive. If the configuration
 is O-He-A, then that means that the nuclear force on He is larger than the electrostatic force from A. Again, a net attraction between OHe and A results.
 
 To settle this classical picture of permanent attraction in the OHe-nucleus system, 
 a proper quantum treatment of the problem is needed. On the one hand, 
 we shall see from the following discussion 
 that simple semiclassical and  perturbative descriptions cannot solve the problem of 
 permanent attraction in the OHe-nucleus system.
 On the other hand, a crucial point may be missing 
 in such  
 treatments:  the correct description of the nuclear effects when 
 the nucleus is close to helium and when neither semiclassical nor perturbative
 approaches are valid. 

\section{Semiclassical model}
The quantum problem involves very
different scales: the OHe binding is of the order of one MeV, and we are
looking for bound states of about one keV. To obtain both in the same framework would
imply
a solution of the 3-body problem at better than one per thousand, which is clearly very
hard.

Fortunately, for the very excited bound states, one can use a simplified method. For
these states, the OHe atom will not dissociate, so we can treat that system as a whole,
allowing a small polarisation in the A direction. Furthermore,
the interaction potential OHe-nucleus can be taken as radial, as the polarisaion of OHe
will be in the A direction. Hence we can use
 spherical coordinates, with the O$^{--}$ fixed at the origin and
  $\overrightarrow{r}$
 the position of the center of the nucleus A. We know in this case that the solutions
of the Schr\"odinger equation take the form $\psi_{k,l,m}(r,\theta,\varphi)=\frac{u_{k,l}
(r)}{r}Y_{l}^{m}(\theta,\varphi)$
where $Y_{l}^{m}(\theta,\varphi)$ are the spherical harmonics and
where the radial part $u_{k,l}(r)$ has to satisfy the radial Schr\"odinger
equation \begin{equation}
\frac{d^{2}u_{k,l}(r)}{dr^{2}}+2m_{A}\left[E_{k,l}-V(r)-\frac{l\left(l+1\right)}
{2m_{A}r^{2}}\right]u_{k,l}(r)=0\label{eq:2b}\end{equation}
 where $l$ is the relative angular momentum, $E_{k,l}$ is the
total energy in the center-of-mass (O$^{--}$) frame, and $V(r)$ is the sum of the nuclear
and of the electrostatic potentials between
OHe and A.

The next simplification comes from the fact that one is looking for weakly bound states,
for which the
WKB method applies, and considerably simplifies the solution.
Finally, we further simplify the problem by approximating the He wave function in the
OHe bound state by a $1s$ hydrogenoid wave function.

The A nucleus is seen as a uniformly charged sphere of charge $Z_{A}$ and of radius
$R_{A}$(fm)$=1.35A^{1/3}$\cite{key-1},
where $A$ is the number of nucleons in the nucleus. Its mass $m_{A}$
is corrected by the nuclear binding energy $B$ given by the Bethe-Weizs\"acker
formula: $m_{A}=Z_{A}m_{p}+N_{A}m_{n}-B$, where $N_{A}$
is the number of neutrons, and $m_{p}$ and $m_{n}$
are the masses of the proton and of the neutron respectively.

The interactions between the OHe atom and the nucleus take two forms: nuclear attraction 
between the helium and the A nucleus at distances $r\lesssim R_{A}$ and electrostatic 
interaction
due to the electrical charges of the components at distances $r\gtrsim R_{A}$.

Out of the nuclear region, the electrostatic interaction is dominant
and can be separated into two contributions : 1) the electrostatic
interaction between the spherical charge distribution of the OHe atom
in its ground state and the spherical charge distribution of the nucleus;
2) the electrostatic interaction between the polarized OHe atom
and the nucleus due to the Stark effect. Therefore, we can write \begin{equation}
V_{Elec}=V_{Coul}+V_{Stark}\label{eq:3b}\end{equation}
 where $V_{Coul}$ corresponds to Coulomb attraction between O$^{--}$
screened by the helium charge distribution and A, and $V_{Stark}$
represents the interaction term of the charged nucleus and the dipole.

Outside the nucleus, i.e. for $r\geqslant R_{A}$, we find for the
Coulomb term 
\begin{eqnarray}
V_{Coul}(r) & = & \frac{3}{8}\left(\frac{-Z_{O}Z_{A}\alpha}{\rho^3
r}\right)e^{-2r/r_{0}}\left[e^{-2\rho}\left\{ \rho^{2}+\frac{5}{4}+\frac{5}{2}\rho
+\left(\frac{1}{2}+\rho\right)\frac{r}{r_0}\right\} \right.\nonumber\\
 &   +&\left.e^{2\rho}\left\{ -\rho^{2}-\frac{5}{4}+\frac{5}{2}\rho
 +\left(-\frac{1}{2}+\rho\right)\frac{r}{r_0}\right\}
 \right]\label{eq:4}
 \end{eqnarray}
 with  the Bohr radius of the OHe atom $r_{0}=1/(m_{He}Z_{O}Z_{He}\alpha)\simeq 1.81$ fm 
 and $\rho=R_A/r_0$.
 This expression can be considered as an improvement of the form
from \cite{key-3} where the nucleus was assumed to be a point-like
particle.

For the Stark potential, we use the formula for the quadratic effect in a constant
electric field \cite{key-4}, taken to be the field of the nucleus at the position of O.
The dipole moment of
the OHe atom in its perturbed ground state  can then be written: \begin{equation}
<q\overrightarrow{R}>=\frac{9}{2}r_{0}^{3}\overrightarrow{E}\label{eq:5b}\end{equation}
 so that, for $r\geqslant R_{A}$, \begin{equation}
V_{Stark}=-q\overrightarrow{R}.\overrightarrow{E}=-\frac{9}{2}r_{0}^{3}E^{2}=-\frac{9}
{2}r_{0}^{3}\frac{Z_{A}^{2}\alpha}{r^{4}}.\label{eq:6b}\end{equation}
Expressions \eqref{eq:4} and \eqref{eq:6b} are valid
when the nuclear effects are negligible.

\begin{figure}
\begin{centering}
\includegraphics[scale=0.7]{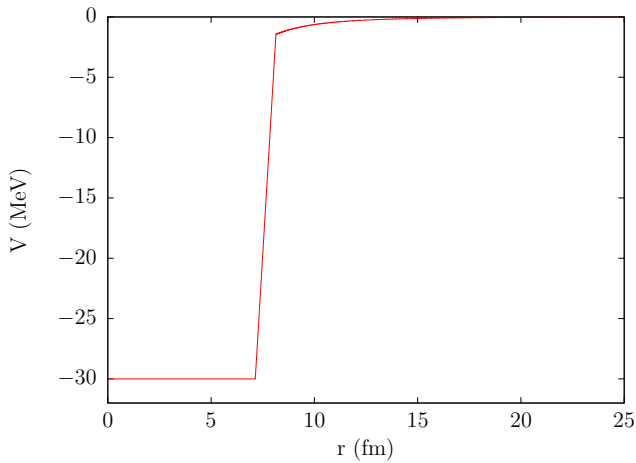}
\par\end{centering}
\caption{Shape of the interaction potential OHe-nucleus, according to Eqs.~\ref{eq:3b} and 
\ref{eq:7}. \label{fig:Shape-of-the}}
\end{figure}

In the nuclear region, we take a trapezoidal nuclear well, which will simplify the WKB 
solution: \begin{equation}
\begin{array}{cccc}
V_{nucl} & = & -V_{0}&{\rm for\ }r\leq R_{*}\\
 & = & \frac{V_{Elec}(R_{*}+2a)+V_{0}}{a}\left(r-R_{*}\right)-V_{0}&{\rm for\ }R_{*}\leq 
 r\leq R_{*}+2a\\
 & = & 0&{\rm for\ }r>R_{*}+2a\end{array}\label{eq:7}\end{equation}
characterized by its depth $V_{0}$ and its diffuseness parameter
$a$ representing the region of $r$ in which it goes linearly from
$-V_{0}$ to $V_{Elec}(R_{*}+a)$. From diffusion experiments of $\alpha$ particles
on nuclei \cite{key-1}, one gets $V_{0}\approx 30$ MeV
for nuclei with $Z_{A}\leq  25$ and $V_{0}=45$ MeV for $Z_{A}>25$,
as well as $a=0.5$ fm. In the following, we shall avoid the transition region
$Z_A\in [21,29]$, which does not contain any nucleus used for direct dark matter
detection. The nuclear radius parameter $R_{*}=(R_{A}+1.3+r_{0})$ fm,
where $R_{A}+1.3$~fm is taken from \cite{key-1} to take the finite
size of the alpha particles into account.
Fig.~\ref{fig:Shape-of-the} shows the form of the potential between OHe and Na.

To find the spectrum corresponding to the potential of Fig.~\ref{fig:Shape-of-the},
we use the approximate WKB solutions, which, once applied to each region,
give a quantization condition for the energy.

For $l=0$, we obtain \begin{equation}
\rho=\left(k-\frac{1}{4}\right)\pi,\,\,\,\, k=1,2,3,...\label{eq:8}\end{equation}
 where $\rho=\int_{0}^{b}k(r)dr$, $b$ is the turning point such that $E=V(b)$
and $k(r)=$\break $\sqrt{2m_{A}\left(E-V(r)\right)}$.

At $l\neq0,$ we know from \cite{key-5} that the behaviour of the
effective potential $V_{eff}(r)=V(r)+\frac{l\left(l+1\right)}{2m_{A}r^{2}}$
at the origin requires to modify the WKB method by applying its solutions
to $u(r)$ after having changed $l\left(l+1\right)$ to $\left(l+\frac{1}{2}\right)^{2}$
in the radial equation : $V_{eff}(r)\rightarrow\widetilde{V}_{eff}(r)=
V(r)+\frac{\left(l+\frac{1}{2}\right)^{2}}{2m_{A}r^{2}}$.
Therefore, the quantization condition becomes \begin{equation}
2e^{2\sigma}\cos\rho+\sin\rho=0\label{eq:9}\end{equation}
 where $\rho=\int_{a}^{b}\widetilde{k}(r)dr$, $\sigma=\int_{0}^{a}
 \widetilde{\kappa}(r)dr$,
$a$ and $b$ are the turning points such that $E=\widetilde{V}_{eff}(a)=\widetilde{V}_{eff}(b)$
and $\widetilde{k}(r)=\sqrt{2m_{A}\left(E-\widetilde{V}_{eff}(r)\right)}$,
$\widetilde{\kappa}(r)=\sqrt{2m_{A}\left(\widetilde{V}_{eff}(r)-E\right)}$.

\subsection{Spectra from a screened Coulomb potential}
\begin{figure}
\begin{centering}
\includegraphics[scale=0.4,angle=90]{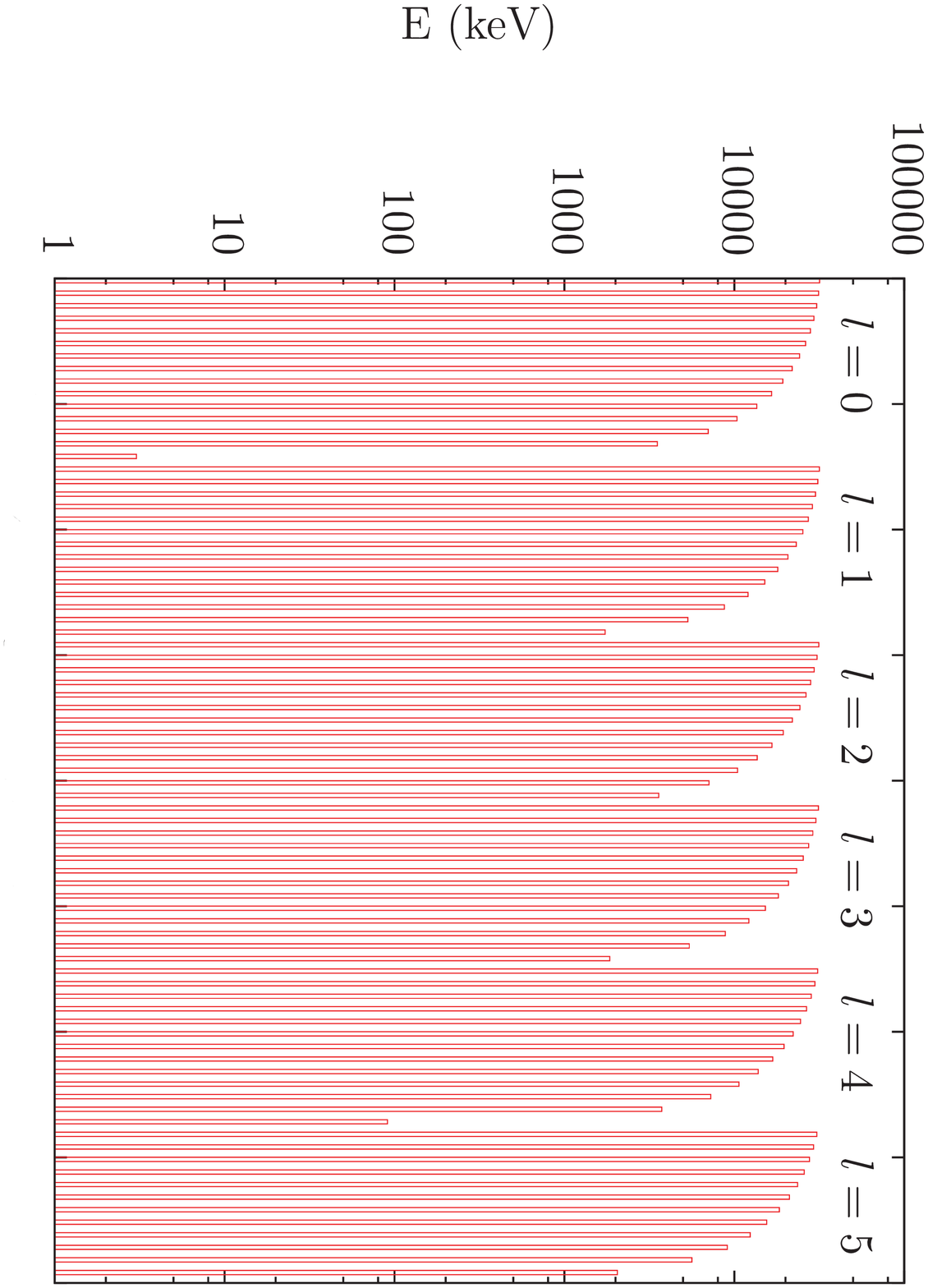}
\par\end{centering}
\caption{Spectrum of the OHe-$^{23}$Na system at different values of $l$.
The energies are in absolute value. $V_{0}=31.9$~MeV and $a=0.5$ fm.}
\label{Flo:Figure 2}
\end{figure}

In \cite{spectro}, the spectrum was considered for a screened Coulomb potential at long
distance, as in Eq.~\ref{eq:4}. We reanalyse this question with our WKB formalism.
For small nuclei $Z_{A}\leq 20$, we first fix the exact value of $V_{0}$ to
obtain the highest level at $-3$ keV for $^{23}$Na from DAMA for $l=0$.
We obtain $V_{0}=31.9$ MeV, in good agreement with \cite{key-1} and use this value
for all nuclei
with $Z_{A}\leq 20$. The spectrum of the OHe-$^{23}$Na system
is shown in Fig.~\ref{Flo:Figure 2}.

\begin{figure}
\begin{centering}
\includegraphics[scale=0.7]{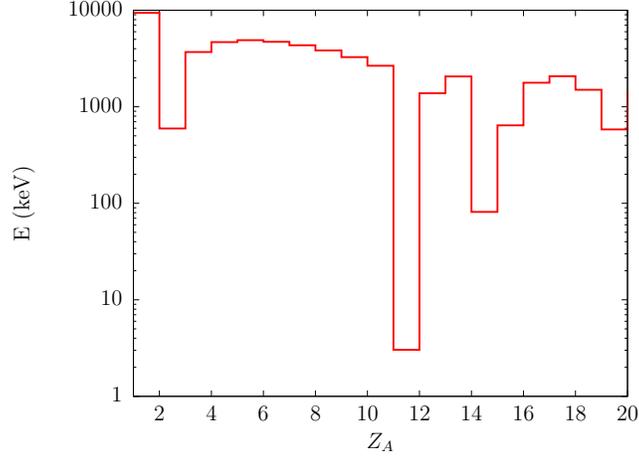}
\par\end{centering}
\caption{Highest-energy level at $l=0$ for stable nuclei from $Z_{A}=1$ to $Z_{A}=20$.
The energies are in absolute value. $V_{0}=31.9$~MeV and $a=0.5$~fm.}
\label{Flo:Figure 3}
\end{figure}

We see a rich spectrum with many levels in the MeV region, corresponding
to nuclear levels, for which the WKB approximation may not be appropriate.
The only level in the keV region is at $l=0$. It can be considered as being
due to the presence of the electrostatic potential. It is remarkable that
other nuclei such that $Z_{A}\leq 20$ do not have keV bound states.
Fig.~\ref{Flo:Figure 3} shows the highest level at $l=0$ for the
most stable nuclei for $Z_{A}$ going from $1$ to $20$. It turns
out that only $^{23}$Na at $Z_{Na}=11$ has a level in the keV
region.

\begin{figure}
\begin{centering}
\includegraphics[scale=0.4,angle=90]{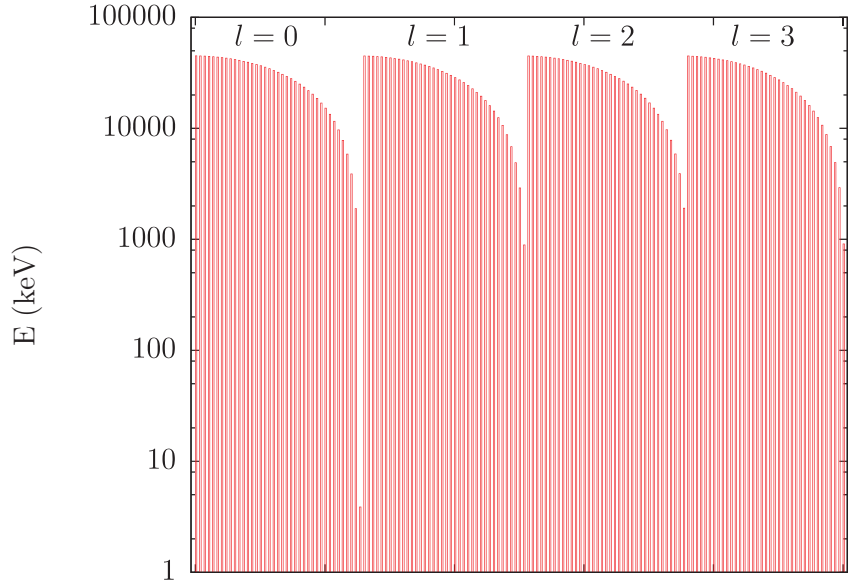}
\par\end{centering}
\caption{Spectrum of the OHe-$^{74}$Ge system at different values of $l$.
The energies are in absolute value. $V_{0}=45$~MeV and $a=0.5$ fm}
\label{Flo:Figure 4}
\end{figure}

For large nuclei $Z_{A}\geq 30$, the data indicate that the nuclear well is deeper.
In this case, we take as a reference germanium $Z_{Ge}=32,\, A=74$ from
CoGeNT, for which we find a highest level at $l=0$ in the keV region
for $V_{0}=45$ MeV, which is precisely the central value from \cite{key-1}
for larger nuclei. This value is used for all nuclei with $Z_{A}\geq 30$.
Fig.~\ref{Flo:Figure 4} represents the
spectrum of the OHe-$^{74}$Ge system. It is of the same kind as for
$^{23}$Na, with only one level in the keV region.

The second column of Table \ref{HighestLargeZ} gives the highest-energy level at $l=0$
for the large
stable nuclei involved in the experiments of interest. According to this model, iodine
and
thallium from DAMA each admit one level in the keV region, while xenon from XENON100
doesn't.

\begin{table}
\begin{center}
\begin{tabular}{|c|c|c|}
\hline
Nuclei  & E(keV) for a screened  & E(keV) for a screened Coulomb\\
  & Coulomb potential  &  potential added to a Stark potential\\
\hline
$^{74}$Ge      &$3.88$ &$1.16$ \\
$^{127}$I      &$0.500$ &$2.31$ \\
$^{132}$Xe     &$540.$ &$2.33$ \\
$^{184}$W      &$350.$ &$1.86$ \\
$^{201}$Tl     &$15.6$&$52.7$ \\
\hline
\end{tabular}
\caption{Highest-energy level at $l=0$ for some heavy stable nuclei from the experiments 
of interest, when $V_{Elec}=V_{Coul}$ (second column), or $V_{Elec}=V_{Coul}+V_{Stark}$ 
(third column). The energies are in absolute value. $V_{0}=45$~MeV and $a=0.5$~fm.
}\label{HighestLargeZ}
\end{center}
\end{table}

\subsubsection{Spectra from a screened Coulomb potential and a Stark potential}
\begin{figure}
\begin{centering}
\includegraphics[scale=0.4,angle=90]{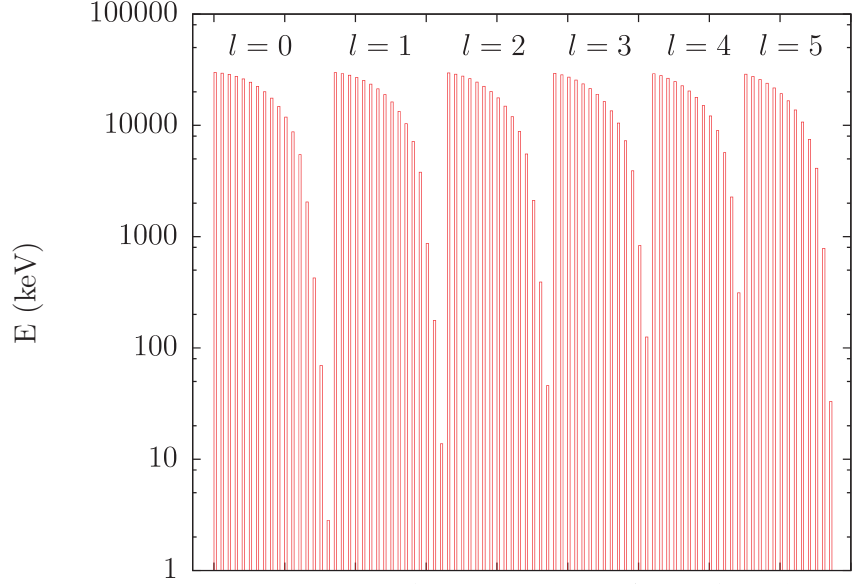}
\par\end{centering}
\caption{Spectrum of the OHe-$^{23}$Na system at different values of $l$,
when $V_{Elec}=V_{Stark}+V_{Coul}$. The energies are in absolute value. $V_{0}=30$~MeV
and $a=0.5$~fm. \label{fig:Figure 6}}
\end{figure}

The results can de discussed in the same way when $V_{Stark}+V_{Coul}$ is
used in the calculations, and the values of $V_{0}$ are identical to the central
experimental values, i.e. $30$ and $45$ MeV, for small and large nuclei respectively.
Fig.~\ref{fig:Figure 6} illustrates the results in the particular case of the 
OHe-$^{23}$Na system.

\begin{figure}
\begin{centering}
\includegraphics[scale=0.7]{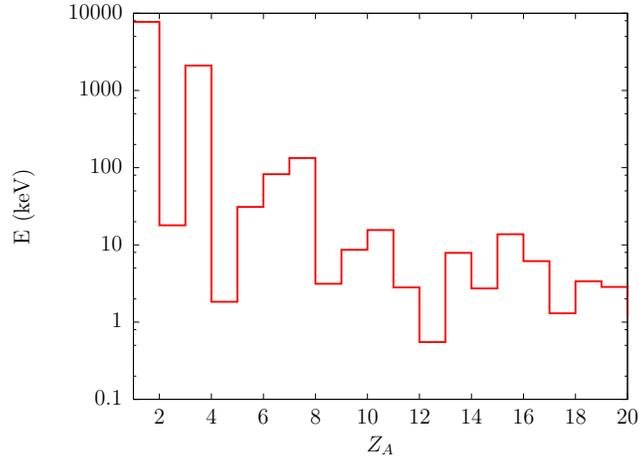}
\par\end{centering}
\caption{Highest level at $l=0$ for stable nuclei from $Z_{A}=1$ to $Z_{A}=20$ due to an 
electromagnetic potential $V_{Stark}+V_{Coul}$. The energies are in absolute value.  
$V_{0}=30$~MeV and $a=0.5$~fm.}
\label{Flo:Figure 6b}
\end{figure}

The major difference lies in the fact that,
in this case, the levels in the keV region are obtained more easily,
with sometimes several keV levels for the same nucleus, especially
for large nuclei. The reason lies in the shape of $V_{Elec}$,
that is deeper and less steep when $V_{Stark}$ is used. Fig.~\ref{Flo:Figure 6b},
as well as third column of Table \ref{HighestLargeZ}, show that most nuclei now have keV
bound states.
Hence the inclusion of the Stark potential seems to destroy the previous interpretation
of the data,
which relied on Na and Ge to be very special nuclei.
\section{Perturbative analysis}
The three-body OHe-nucleus bound-state problem can be simplified in another way,
by noting that helium is much lighter than the A nuclei. Given this, one can simplify the 
total hamiltonian
of the system, written in the reference frame of the O$^{--}$ particle, and choosing the
$z$ axis in the direction of A
to: \begin{equation}
H\approx -\frac{1}
{2m_{He}}\triangle_{1}+V_{OHe}\left(r_{1}\right)+V_{OA}\left(R\right)+V_{HeA}\left(r_{12}\right)\label{eq:1}\end{equation}
 in which the kinetic energy term of the A nucleus has been neglected
and where $\overrightarrow{r}_{1}$ is the position of the He nucleus,
$R$ is the distance of the A nucleus on the positive part of the
$z-$axis and $r_{12}$ is the distance between He and the
A nucleus. $V_{IJ}$ stand for the interaction
potential between I and J.
We are thus left with the one-body
problem of He in a total potential depending on the parameter
$R$.
We shall consider here the contribution of the
external A nucleus as a perturbation to the OHe atom. The hamiltonian \eqref{eq:1}
can be rewritten
as the sum of an unperturbed part $H_{0}$ and a perturbation $W$:
 \begin{equation}
H=H_{0}+W\label{eq:2c}\end{equation}
 where $H_{0}=-\frac{1}{2m_{He}}\triangle_{1}+V_{OHe}\left(r_{1}\right)$
corresponds to the isolated OHe atom and where 
$W=V_{OA}\left(R\right)+V_{HeA}\left(r_{12}\right)$
is due to the presence of the external nucleus.
We use here the following two-body interaction potentials :
\[
\begin{array}{llll}
V_{OHe}(r_{1}) & = & -\frac{Z_{O}Z_{He}\alpha}{r_{1}}&\\~\\
V_{OA}(R) & = & -\frac{Z_{O}Z_{A}\alpha}{R}&,\,R>R_{A}\\
 & = & -\frac{Z_{O}Z_{A}\alpha}{2R_{A}}\left(3-\frac{R^{2}}{R_{A}^{2}}\right)& ,
 \,R<R_{A}\\~\\
V_{HeA}(r_{12}) & = & \frac{Z_{He}Z_{A}\alpha}{r_{12}}+\frac{-V_{0}}{1+e^{\left(r_{12}-
R_{A}\right)/a}}&,\,r_{12}>R_{A}\\
 & = & \frac{Z_{He}Z_{A}\alpha}{2R_{A}}\left(3-\frac{r_{12}^{2}}
 {R_{A}^{2}}\right)+\frac{-V_{0}}{1+e^{\left(r_{12}-R_{A}\right)/a}}& ,\, 
 r_{12}<R_{A}\end{array}\]
where, in order to simplify the calculation,
the OHe atom is treated as a hydrogenoid system,
and where the A nucleus is seen as a sphere of radius $R_{A}$(fm)$=1.35\times A^{1/3}$
\cite{key-1}
and charge $Z_{A}$ in the definitions of potentials $V_{OA}$ and
$V_{HeA}$ that are therefore point (O$^{--}$ or He$^{++}$) - sphere
(A) interaction potentials. A Woods-Saxon potential of parameters
$V_{0}$ and $a$ has been added in $V_{HeA}$ to take the nuclear
interaction of both nuclei into account.

We are studying the perturbed ground state $E_{0}(R)$ of the OHe atom
under the influence of the external perturbation $W(R)$. The perturbed
energy is therefore an approximation of the energy of the total O-He-A
system, described by the hamiltonian \eqref{eq:2c}. If this energy presents
a minimum for some $R_{b}$, then the system will tend to this configuration
to minimize its energy, and we will get a stable OHe-nucleus bound
state of length $R_{b}$ and energy $E_{0}(R_{b})$. If there
is no minimum,
then we will have to conclude that no stable bound state can form
at those distances.

In the following, we shall go to 3rd-order perturbation theory.
We assume that $H_0$ has a spectrum $|\psi_{n}^{0}>$ of eigenfunctions with eigenvalues
$E_n^0$, and we assume that the unperturbed energy level is non-degenerate.
The formulae for the wave function at order 2 and for the energy at order 3 are given by:

\begin{eqnarray}
E_{n} & = & E_{n}^{0}\nonumber\\
&+&<\psi_{n}^{0}|W|\psi_{n}^{0}>\nonumber\\
&+& \sum_{i,{p\neq n}}\frac{|<\psi_{p,i}^{0}|W|\psi_{n}^{0}>|^{2}}{E_{n}^{0}-
E_{p}^{0}}\nonumber\\
&-& <\psi_{n}^{0}|W|\psi_{n}^{0}>\sum_{i,{p\neq n}}\frac{|<\psi_{p,i}^{0}|W|
\psi_{n}^{0}>|^{2}}{(E_{n}^{0}-E_{p}^{0})^{2}}\\
&+& \sum_{i,{p\neq n}}\sum_{i',{p'\neq n}}\frac{<\psi_{p',i'}^{0}|W|
\psi_{n}^{0}><\psi_{p,i}^{0}|W|\psi_{p',i'}^{0}><\psi_{n}^{0}|W|\psi_{p,i}^{0}>}
{\left(E_{n}^{0}-E_{p}^{0}\right)\left(E_{n}^{0}-E_{p^{,}}^{0}\right)},\nonumber
\end{eqnarray}
\begin{eqnarray}
|\psi_{n}> & = & |\psi_{n}^{0}>\nonumber\\
&+&\sum_{i,p\neq n}\frac{<\psi_{p,i}^{0}|W|\psi_{n}^{0}>}{E_{n}^{0}-E_{p}^{0}}|
\psi_{p,i}^{0}> \nonumber\\
&-&<\psi_{n}^{0}|W|\psi_{n}^{0}>\sum_{i,{p\neq n}}\frac{<\psi_{p,i}^{0}|W|
\psi_{n}^{0}>}{(E_{n}^{0}-E_{p}^{0})^{2}}|\psi_{p,i}^{0}>\nonumber\\
&-&\frac{1}{2}\sum_{i,{p\neq n}}\frac{|<\psi_{p,i}^{0}|W|\psi_{n}^{0}>|^{2}}
{(E_{n}^{0}-E_{p}^{0})^{2}}|\psi_{n}^{0}>\nonumber\\
&+& \sum_{i,{p\neq n}}\sum_{i',{p'\neq n}}
\frac{<\psi_{p',i'}^{0}|W|\psi_{n}^{0}><\psi_{p,i}^{0}|W|\psi_{p',i'}^{0}>}
{\left(E_{n}^{0}-E_{p}^{0}\right)\left(E_{n}^{0}-E_{p'}^{0}\right)}|
\psi_{p,i}>.\end{eqnarray}

In our case, the non-degenerate unperturbed energy $E_{n}^{0}$
is the ground level of the hydrogenoid OHe atom: \begin{equation}
E_{n}^{0}=E_{OHe}=-\frac{1}{2}m_{He}\left(Z_{O}Z_{He}\alpha\right)^{2}\simeq-1.58\, {\rm MeV}\end{equation}
 and the unperturbed eigenfunction $|\psi^{0}_{p,i}>$ are those of
 the hydrogen atom: 
 \begin{equation}
\psi^{0}_{p,i}(\overrightarrow{r_{1}})=\psi^{0}_{n,l,m}(\overrightarrow{r_{1}})=
R_{n,l}(r_{1})Y_{l}^{m}(\theta_{1},\varphi_{1}),
\label{eq:17}
\end{equation}
where the $Y_l^m$ are the normalised spherical harmonics, and where the radial part 
$R_{n,l}$ is given by \begin{equation}
R_{n,l}(r_{1})=C_{n,l}\times r_{1}^{l}\sum_{q=0}^{n-l-1}c_{q}\left(\frac{r_{1}}
{r_{0}}\right)^{q}e^{-\frac{r_{1}}{n r_{0}}},\label{eq:18}\end{equation}
 $C_{n,l}$ being the normalization coefficient of $R_{n,l}$ and
$r_{0}$ being the Bohr radius of the OHe atom.
The coefficients $c_{q}$ in \eqref{eq:18} are recursively given
by $\frac{c_{q}}{c_{q-1}}=-\frac{2\left(1-\frac{q+l}{n}\right)}{q\left(q+2l+1\right)}$.

\subsection{Correction to the OHe energy}
\begin{figure}
\begin{centering}
\includegraphics[scale=0.7]{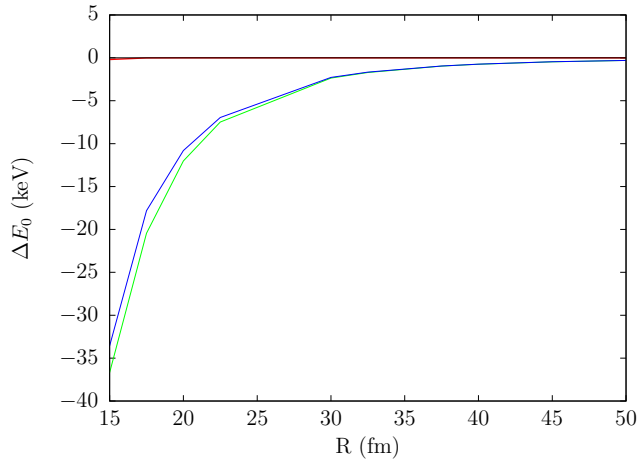}
\par\end{centering}
\caption{$\triangle E_{0}=E_{0}(R)-E_{OHe}$ (keV) up to orders $1$ (upper),
$2$ (lower) and $3$ (middle) for an external sodium nucleus, as a
function of its distance $R$ (fm). $V_{0}=30$ MeV and $a=0.5$~fm.
\label{fig:-up-to}}
\end{figure}

First, we consider the effect of an approaching sodium nucleus on the
OHe energy. Fig.~\ref{fig:-up-to} shows the results
for $\triangle E_{0}=E_{0}(R)-E_{OHe}$ for $V_{0}=30$~MeV and $a=0.5$ fm. We see that 
order $1$
doesn't bring a large modification to the unperturbed energy ($\sim10^{-3}$ keV
for $R$ between $50$ and $15$~fm), while order $2$ gives the
largest correction ($\sim1-10$ keV for $R$ between $50$ and $15$~fm).
This change from order $1$ to order $2$ justifies the inclusion
of order $3$ in the calculations, but it turns out that this one
doesn't modify greatly the results from order $2$, and that is why
order $4$ has not been added. We see on Fig.~\ref{fig:-up-to}
that $\triangle E_{0}$ is always decreasing, in other words that
there is no minimum in this curve in the region of validity of the
perturbative calculation.

\begin{figure}
\begin{centering}
\includegraphics[scale=0.7]{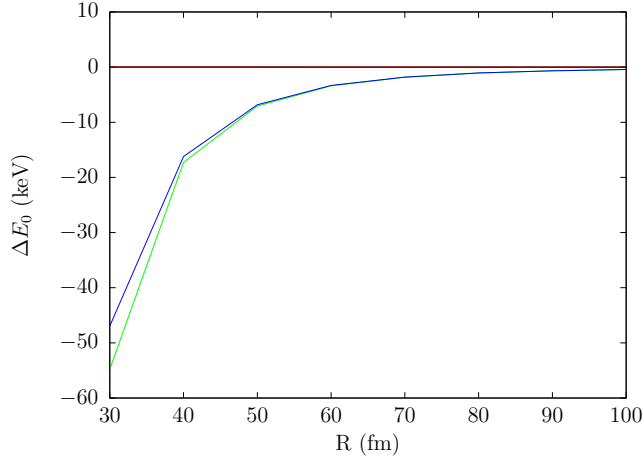}
\par\end{centering}
\caption{$\triangle E_{0}=E_{0}(R)-E_{OHe}$ (keV) up to orders $1$ (upper),
$2$ (lower) and $3$ (middle) for an external iodine nucleus, as a
function of its distance $R$ (fm). $V_{0}=45$ MeV and $a=0.5$ fm.\label{fig:-up-to3}}
\end{figure}

Similar results hold if one strengthens the nuclear potential, or if one considers 
different nuclei, as shown in Fig.~\ref{fig:-up-to3} in the case of iodine.

\subsection{Interaction with the incoming nucleus and polarization}
In the same way, we can calculate the electrostatic and nuclear interaction
energies between the perturbed charge distribution of the helium nucleus
and the charge distribution of the A nucleus as a function of its
distance $R$, as well as the mean position of He
along the $z-$axis, that is the polarization of the OHe under the
influence of the external nucleus.

The electrostatic interaction energy between two
charge distributions is given by \begin{equation}
E_{el}=\int_{V1}\int_{V2}\frac{\rho_{1}\left(\overrightarrow{r_{1}}\right)\rho_{2}\left(\overrightarrow{r_{2}}\right)}
{\left|
\overrightarrow{r_{1}}-\overrightarrow{r_{2}}\right|}d\overrightarrow{r_{1}}d\overrightarrow{r_{2}}\label{eq:20}\end{equation}
 where each integral is performed over the extension $V_{1}$ or $V_{2}$
of the corresponding charge distribution and where $\rho_{1}$ and
$\rho_{2}$ are the charge densities of each distribution.
In the following, we take the first-order version of $\psi$ for He (as it is responsible 
for the dominant second-order shift in energy), and 
$\rho_{1}\left(\overrightarrow{r_{1}}\right)=|
\psi_{He}\left(\overrightarrow{r_{1}}\right)|^2$, while the
 external nucleus is treated as a uniform sphere.

\begin{figure}
\begin{centering}
\includegraphics[scale=0.7]{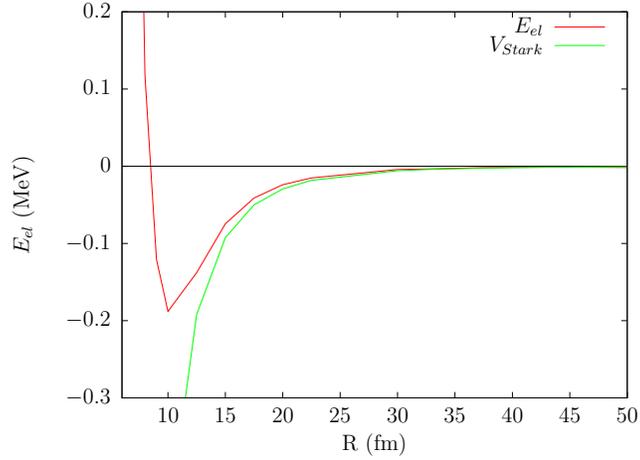}
\par\end{centering}
\caption{Electrostatic interaction energy $E_{el}$ (MeV) at order $1$ (upper)
compared to the Stark interaction energy $V_{Stark}$ (MeV) (lower)
as a function of the distance $R$ (fm) of an external sodium nucleus;
$V_{0}=30$ MeV and $a=0.5$ fm. \label{fig:Electrostatic-interaction-energy}}
\end{figure}

The results are compared to the Stark
potential used in the previous section in
Fig.~\ref{fig:Electrostatic-interaction-energy}.
We see, as might be expected, that the simplifying
assumption of the constant electrical field for the nucleus is reasonable
at large distance, while the gap becomes more pronounced around $15$ fm,
because the uniform Coulomb field is always
stronger than the true one. The fact that $E_{el}$
becomes repulsive at shorter distance is due to the change of the
polarization of the OHe atom under the influence of the nuclear force
of the sodium nucleus, which makes the helium component turning to
positive mean $z_{1}$, that is, towards the external nucleus.

\begin{figure}
\begin{centering}
\includegraphics[scale=0.7]{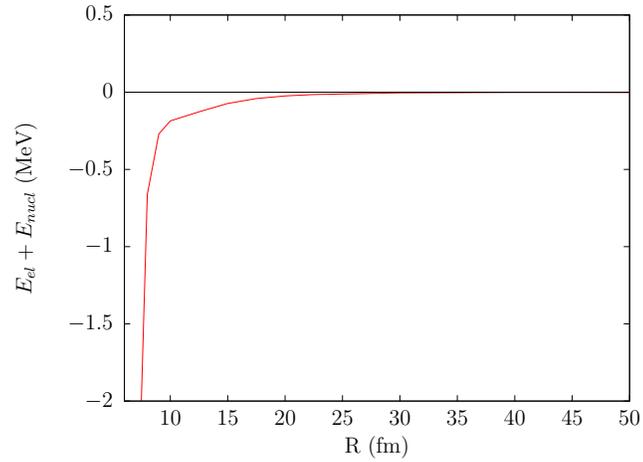}
\par\end{centering}
\caption{Total interaction energy $E_{el}+E_{nucl}$ (MeV) at order $1$ as
a function of the distance $R$ (fm) of an external sodium nucleus;
$V_{0}=30$~MeV and $a=0.5$~fm. \label{fig:Total-interaction-energy} }
\end{figure}

We can also integrate the Woods-Saxon potential $\frac{-V_{0}}{1+e^{\left(r-
R_{A}\right)/a}}$
over the distribution of the helium nucleus to get the total nuclear interaction energy 
$E_{nucl}$. Adding it to the previous contribution gives us the curve of 
Fig.~\ref{fig:Total-interaction-energy}, which has no sign of a potential barrier.

\begin{figure}
\begin{centering}
\includegraphics[scale=0.7]{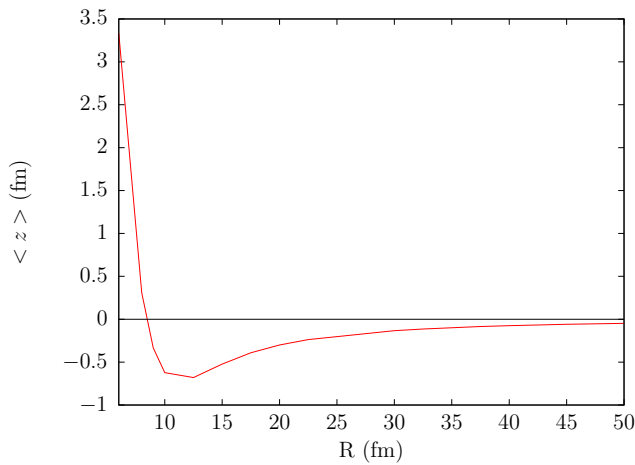}
\par\end{centering}
\caption{Polarization $<z>$ (fm) at order $1$ as a function of the distance
$R$ (fm) of an external sodium nucleus; $V_{0}=30$ MeV and $a=0.5$ fm.
\label{fig:Polarization--at}}
\end{figure}

Finally, we can calculate the  mean value of the position $z_{1}$ of the helium nucleus 
along
the $z-$axis, that is the polarization of the OHe atom, which is simply
obtained by \begin{equation}
<z>=<\psi_{He}|z_{1}|\psi_{He}>=\int d\overrightarrow{r_{1}}\, z_{1}|
\psi_{He}\left(\overrightarrow{r_{1}}\right)|^{2}\label{eq:25}\end{equation}
 Fig.~\ref{fig:Polarization--at} represents the evolution of this polarization
as a function of $R$. It can be seen that, at large
distance, the polarization is negative due to Coulomb repulsion
between nuclei, as expected. Thus, Fig.~\ref{fig:Polarization--at}
shows that the OHe atom gets polarized, for $R\lesssim10$ fm, in
a direction that could allow repulsion, provided that the nuclear
force is not already too strong at such distance. The addition of
the nuclear interaction with $V_{0}=30$ MeV and $a=0.5$~fm in Fig. 
\ref{fig:Total-interaction-energy} shows
that this condition is in fact not satisfied, giving rise to an attractive
force at all distances. The modification of the nuclear parameters
$V_{0}$ and $a$ (for example $V_{0}=10,\,100,\,200$~MeV, $a=0.5,\,1.5$ fm),
as well as the external nucleus, doesn't radically change the results,
modifying only the distance from which the potential falls to nuclear
values.

\section{Conclusion}
The advantages of the OHe composite-dark-matter scenario is that it is minimally related
to the parameters of new physics and is dominantly based on the effects of known atomic
and nuclear physics.
However, the proper quantum treatment of this problem turns out to be rather
complicated and involves several open questions.

We have presented here the state of the art of our studies of the nuclear physics of the
OHe atoms, and found a difficulty in proving the original assumption of a potential
barrier developing between OHe and the nucleus A, both classically and in perturbation 
theory.

Open questions for further analysis nevertheless remain:
\begin{itemize}
\item[(a)] for the distances under consideration the size of the He nucleus may not be
negligible and it may not be sufficient to treat it as a point-like particle;
\item[(b)] beyond the nucleus the nuclear force falls down exponentially but it may be
strong enough to cause a non-homogeneous perturbation of the OHe atomic ground state;
\item[(c)] the nuclear force indeed leads to a change of the OHe polarization that might 
result in the creation of a dipole Coulomb barrier, as shown in the perturbative 
calculation, but this happens when the perturbative approach is no longer valid, and
one should thus solve the Schrödinger equation numerically in this regime.
\end{itemize}
The answer to these open questions may be crucial for asserting the nuclear-physics
basis of the OHe model. If there is no dipole Coulomb barrier between OHe and nucleus,
one gets
a spectrum of states, which could have transitions to each other. Although the spectra
we showed in the third section are not reliable in the nuclear region, it is clear that
$\alpha$ particles will have nuclear bound states. Without a barrier, their transitions
to them will be fast and dramatic.

Hence, the model cannot work if no repulsive interaction appears at some distance between
OHe and the nucleus, and the solution to the open questions of OHe nuclear physics is 
vital for the composite-dark-matter scenario.

\section*{Acknowledgements}
We thank A.G. Mayorov and E. Yu. Soldatov for many discussions at the
beginning of this work, and J. Cugnon and D. Mancusi for sharing their
knowledge of the intricacies of nuclear physics with us.

\end{document}